\documentclass[twocolumn,showkeys,aps,prl,showpacs]{revtex4-1}
\usepackage{graphicx}
\usepackage[CJKbookmarks,dvipdfm,colorlinks,linkcolor=blue,citecolor=blue]{hyperref}

\begin{document}

\title{Thermoelectric properties of  half-Heusler $\mathrm{ZrNiPb}$ by using first principles calculations}

\author{San-Dong Guo}
\affiliation{Department of Physics, School of Sciences, China University of Mining and
Technology, Xuzhou 221116, Jiangsu, China}
\begin{abstract}
We investigate electronic structures and  thermoelectric properties of  recent synthetic half-Heusler $\mathrm{ZrNiPb}$ by using  generalized gradient approximation (GGA) and GGA plus spin-orbit coupling (GGA+SOC).  Calculated results show that $\mathrm{ZrNiPb}$ is a indirect-gap semiconductor. Within the constant scattering time approximation, semi-classic transport coefficients are performed through solving Boltzmann transport equations. It is found that the  SOC has  more  obvious  influence on power factor in p-type doping than in n-type doping, leading to a detrimental effect  in p-type doping. These can be explained  by considering the SOC influences on the valence bands and conduction bands near the Fermi level. The lattice thermal conductivity as a function of temperature is calculated, and the corresponding lattice thermal conductivity is 14.5  $\mathrm{W m^{-1} K^{-1}}$ at room temperature. By comparing the experimental transport coefficients with calculated ones, the
scattering time is attained for 0.333 $\times$ $10^{-14}$ s.  Finally, the thermoelectric figure of merit $ZT$  can be attained, and the $ZT$ value can be as high as 0.30 at high temperature by choosing  appropriate doping level. It is possible to reduce lattice thermal conductivity  by point defects and boundaries, and make  half-Heusler $\mathrm{ZrNiPb}$ become potential candidate for efficient thermoelectricity.

\end{abstract}


\maketitle

\section{Introduction}
The  performance  of thermoelectric material can be  characterized by the dimensionless thermoelectric figure of merit\cite{s1,s2}, $ZT=S^2\sigma T/(\kappa_e+\kappa_L)$, where S, $\sigma$, T, $\kappa_e$ and $\kappa_L$ are the Seebeck coefficient, electrical conductivity, absolute working temperature, the electronic and lattice thermal conductivities, respectively. Various thermoelectric materials have been identified, such as  relatively low temperature operation Bismuth-tellurium systems\cite{s3,s4}, high temperature operation silicon-germanium alloys\cite{s5,s6}, moderate temperature operation lead chalcogenides\cite{s7,s8}.
Heusler compounds  can be applied to  spintronics, superconductors, topological insulators and thermoelectrics\cite{s11}.
Due to being environmentally friendly, mechanically and thermally robust, 18-valence electron half-Heusler has been widely investigated for thermoelectric application\cite{s12,s13,s16,s17}. Many theoretical reports on thermoelectric properties of  half-Heusler focus on electronic part\cite{t5,t51,t52,t53,t54,t55,gsd1,hf1}. Recently, the lattice  thermal conductivities of half-Heusler alloy TiNiSn and ZrCoSb by first principles calculations are attained, and the calculated lattice thermal conductivities are close to the experimental values\cite{t56,t561}.

In ref.\cite{nc}, by using first principles calculations, 54 of the 400 unreported 18-valence electron half-Heuslers  should be
stable, and 15 have been  grown, whose  X-ray experimental results agree with the predicted crystal structures.
$\mathrm{ZrNiPb}$ is synthesized as a single phase, and the room temperature  power factor is as high as  5.2 $\mathrm{\mu W cm^{-1} K^{-2}}$.  However, the thermal conductivity of $\mathrm{ZrNiPb}$ hasn't been measured by the related experiments.
Here, the electronic structures and  thermoelectric properties of  half-Heusler $\mathrm{ZrNiPb}$  are studied by the first
principles calculations and  Boltzmann transport equations within the constant scattering time approximation.   The SOC has been proved to be very important for power factor calculations of many thermoelectric materials\cite{gsd1,gsd2,e1}, so the SOC effect is considered in our calculations.  Calculated results show SOC has larger effects on power factor in p-type doping than in n-type doping, and has a reduced influence for p-type. In this work, the electronic part of thermoelectric properties of  half-Heusler $\mathrm{ZrNiPb}$ is  not only calculated,  but also the lattice
lattice thermal conductivity is attained. The calculated room temperature lattice thermal conductivity is 14.5  $\mathrm{W m^{-1} K^{-1}}$.  Although calculating  scattering time is difficult, it can be attained by the comparison between experimental and calculated  transport coefficients. Finally, the dimensionless thermoelectric figure of merit $ZT$ is calculated, and can be up
to 0.3 at high temperature by the optimized doping.

\begin{figure}
  \includegraphics[width=8cm]{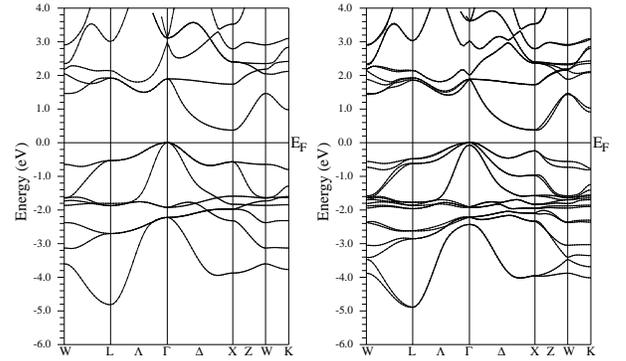}
  \caption{The energy band structures of $\mathrm{ZrNiPb}$ by using GGA (Left) and GGA+SOC (Right).}\label{band}
\end{figure}

\begin{figure}
  \includegraphics[width=7cm]{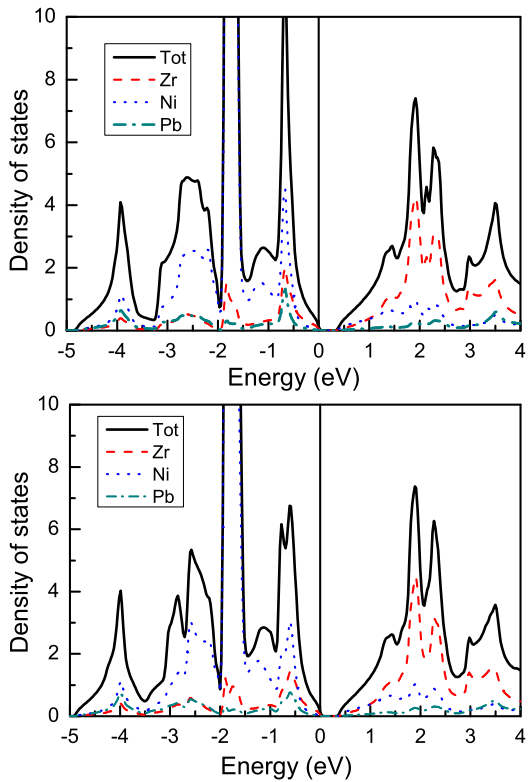}
  \caption{(Color online) The total and partial densities of states  of  $\mathrm{ZrNiPb}$ by using GGA (Top panel) and GGA+SOC (Bottom panel).}\label{dos}
\end{figure}

\begin{figure*}
  \includegraphics[width=15.6cm]{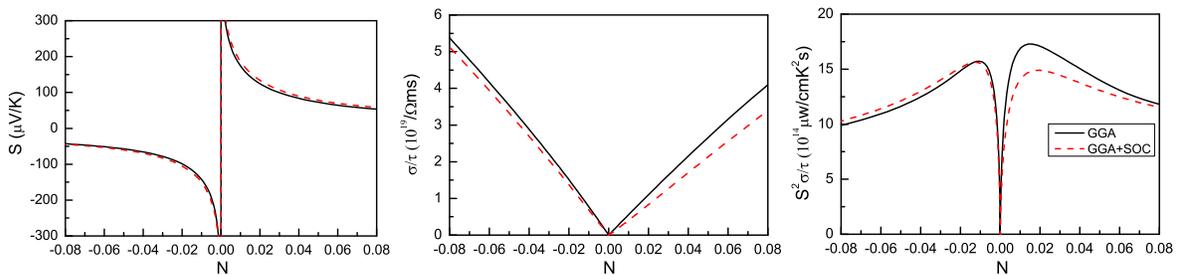}
  \caption{(Color online) At temperature of 300 K,  transport coefficients  as a function of doping levels (electrons [minus value] or holes [positive value] per unit cell):  Seebeck coefficient S (Left),  electrical conductivity with respect to scattering time  $\mathrm{\sigma/\tau}$, (Middle) and power factor with respect to scattering time $\mathrm{S^2\sigma/\tau}$ (Right) calculated with GGA (Solid line) and GGA+SOC (Dotted line). }\label{s0}
\end{figure*}
\begin{figure}
  \includegraphics[width=7cm]{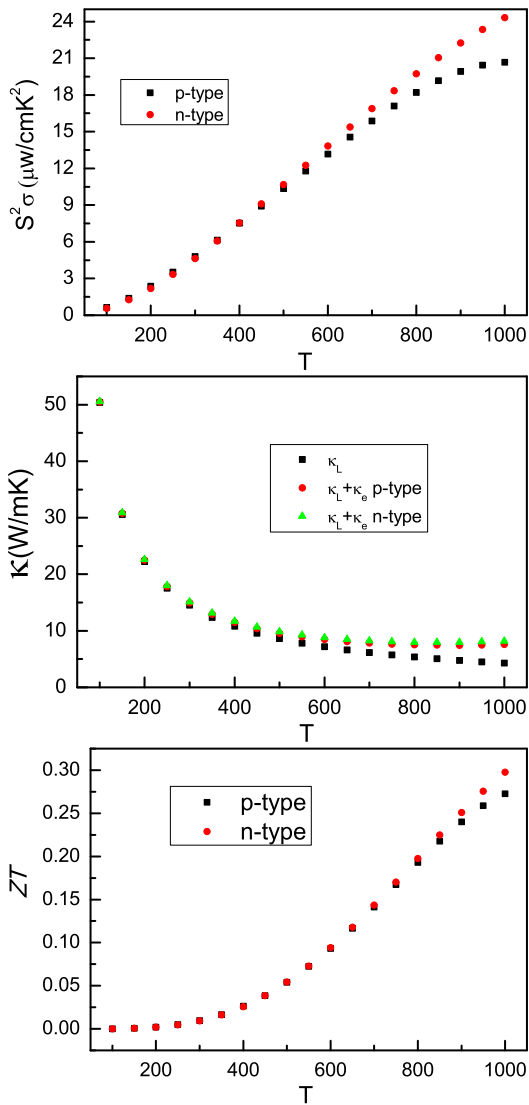}
  \caption{(Color online) The  power factor  $\mathrm{S^2\sigma}$,  thermal conductivity $\mathrm{\kappa}$ (electronic $\mathrm{\kappa_e}$ and lattice  $\mathrm{\kappa_L}$ thermal conductivities) and $ZT$ as a function of temperature with the doping concentration of $\mathrm{5\times10^{20}cm^{-3}}$ for p-type and n-type, and the scattering time $\mathrm{\tau}$  is 0.333 $\times$ $10^{-14}$ s.}\label{s1}
\end{figure}
The rest of the paper is organized as follows. In the next
section, we shall give our computational details. In the third section, we shall present our main calculated results and
analysis. Finally, we shall give our conclusion in the fourth
section.

\section{Computational detail}
We use a full-potential linearized augmented-plane-waves method
within the density functional theory (DFT) \cite{1}, as implemented in
the package WIEN2k \cite{2}.  We use GGA \cite{pbe} to do our main DFT
calculations.  The full relativistic effects are calculated
with the Dirac equations for core states, and the scalar
relativistic approximation is used for valence states
\cite{10,11,12}. The SOC was included self-consistently
by solving the radial Dirac equation for the core electrons
and evaluated by the second-variation method\cite{so}. We use 10000 k-points in the
first Brillouin zone for the self-consistent calculation.
We make harmonic expansion up to $\mathrm{l_{max} =10}$ in each of the atomic spheres, and
set $\mathrm{R_{mt}*k_{max} = 8}$. The self-consistent calculations are
considered to be converged when the integration of the absolute
charge-density difference between the input and output electron
density is less than $0.0001|e|$ per formula unit, where $e$ is
the electron charge. Transport calculations
are performed through solving Boltzmann
transport equations within the constant
scattering time approximation as implemented in
BoltzTrap\cite{b}, which has been applied successfully to several
materials\cite{b1,b2,b3}. To
obtain accurate transport coefficients, we use 200000 k-points in the
first Brillouin zone for the energy band calculation. The  lattice thermal conductivities are calculated
by using Phono3py+VASP codes\cite{pv1,pv2,pv3,pv4}. For the third-order force constants, 2$\times$2$\times$2 supercells
are built, and reciprocal
spaces of the supercells are sampled by  2$\times$2$\times$2 meshes. To compute lattice thermal conductivities, the
reciprocal spaces of the primitive cells  are sampled using the 20$\times$20$\times$20 meshes.

\section{MAIN CALCULATED RESULTS AND ANALYSIS}
The 18-valence electron half-Heusler $\mathrm{ZrNiPb}$ possesses,  forming a MgAgAs type
of structure, space group $F\bar{4}3m$, where Zr, Ni, and Pb atoms occupy
Wyckoff positions 4a (0, 0, 0), 4c (1/4, 1/4,
1/4)  and 4b (1/2, 1/2, 1/2) positions, respectively.
The  optimized lattice constants $a$=6.27 $\mathrm{{\AA}}$ is used to do our DFT calculations, and present energy band structures and density of states (DOS)  by using GGA and GGA+SOC in \autoref{band} and \autoref{dos}.
$\mathrm{ZrNiPb}$ is a indirect-gap semiconductor, with the conduction band minimum (CBM) at high symmetry point X and  valence band maximum (VBM) at the $\Gamma$ point, and the energy band gap value is about 0.37 eV by using both GGA and GGA+SOC.
 The DOS's show that the strong hybridization of d states of Zr and Ni atoms produces the energy band gap.  The bottom of the conduction bands are constructed  mostly  by Zr-d and Ni-d states,  while the valence bands near the Fermi level are dominated by the Zr-d state hybridized with the Ni-d and Pb-p states. The SOC  can  remove the band degeneracy.
 It is clearly seen that SOC leads  to a  remarkable spin-orbital splitting value of 0.73 eV at high symmetry  X point of valence
 bands near the Fermi level, and 0.07 eV at $\Gamma$ point.
 The SOC has  weaker  effect on the conduction bands near the Fermi level than on the valence bands.
Compared to  half-Heusler $\mathrm{ANiB}$ (A=Ti, Hf, Sc, Y; B=Sn, Sb, Bi)\cite{gsd1}, a major difference is that the energy between VBM and the extremum of  valence at X point is more close, which leads to different SOC effects on  Seebeck coefficient.

Here, the band structures are supposed to be independent of the temperature and doping, and   the semi-classic transport coefficients as a function
of doping level are calculated within constant scattering time approximation Boltzmann theory.
\autoref{s0}  shows the Seebeck coefficient S, electrical conductivity with respect to scattering time  $\mathrm{\sigma/\tau}$ and  power factor with respect to scattering time $\mathrm{S^2\sigma/\tau}$  as  a function of doping levels  at the temperature of 300 K by using GGA and GGA+SOC.  The negative doping levels, being related with conduction bands,  imply the
n-type doping with the negative Seebeck coefficient, and
the positive doping levels, being connected with valence bands, mean p-type doping with the positive Seebeck coefficient.

In narrow-gap semiconductors,
the large slope of DOS near the energy gap may
give rise to a large Seebeck coefficient\cite{hf1}. Calculated results show that the slope of DOS  by using GGA+SOC near the energy band gap is larger than that by using GGA  for both valence and conduction bands.
Calculated results show, in both p-type and n-type doping,  SOC has a enhanced  influence on the Seebeck coefficient  S (absolute value). This is different from  half-Heusler $\mathrm{ANiB}$ (A=Ti, Hf, Sc, Y; B=Sn, Sb, Bi)\cite{gsd1}, which is due to closer valence extremum between $\Gamma$ and X points(Band degeneracy, namely  band convergence, can enhance Seebeck coefficient\cite{gsd2,bc}, and here is SOC driven degeneracy.).
However, the SOC has a detrimental influence on the  $\mathrm{\sigma/\tau}$ for both p-type and n-type doping, which can be explained by that SOC leads to more  localized  bands.
When SOC isn't considered,  the best   power factor in p-type doping  is larger than that in n-type. However, including SOC, the best n-type power factor  is larger than the best p-type one in considered doping range. Similar SOC effects on best power factor can be found in $\mathrm{Mg_2Sn}$\cite{gsd2}, which agrees well with experimental results.
In n-type doping, SOC can reduce  the power factor in low doping levels, but can also enhance one in high doping levels.
In p-type doping, SOC has more obvious effect on power factor, leading to a detrimental influence in the considered doping levels. The maximum power factor (MPF) by using GGA+SOC in p-type doping is about 13.8\% smaller than that with GGA. Therefore, it is necessary  to consider SOC  in the theoretical prediction of thermoelectric properties of $\mathrm{ZrNiPb}$.

Calculating  $\tau$ from the first principles is challenging, but it can be attained by comparing the related experimental results.  At room temperature, the measured electron conductivity and Seebeck coefficient of $\mathrm{ZrNiPb}$   are
220.1 $\mathrm{\Omega^{-1} cm^{-1}}$ and -153.9 $\mathrm{\mu V K^{-1}}$, respectively, which gives a power
factor  5.2 $\mathrm{\mu W cm^{-1} K^{-2}}$\cite{nc}.  By comparing  measured and calculated  Seebeck coefficient, the doping level is determined to be -0.0097, and the $\tau$ is attained for 0.333 $\times$ $10^{-14}$ s by  making a comparison between experimental and  theoretical  electron conductivities.  The power factor $\mathrm{S^2\sigma}$ with the doping level  -0.0097 is calculated by using $\tau$=0.333 $\times$ $10^{-14}$ s, and the attained valve is 5.16  $\mathrm{\mu W cm^{-1} K^{-2}}$, being in agreement with experimental value 5.2 $\mathrm{\mu W cm^{-1} K^{-2}}$,  which confirms the reliability of our calculated results.
\begin{figure}
  \includegraphics[width=7cm]{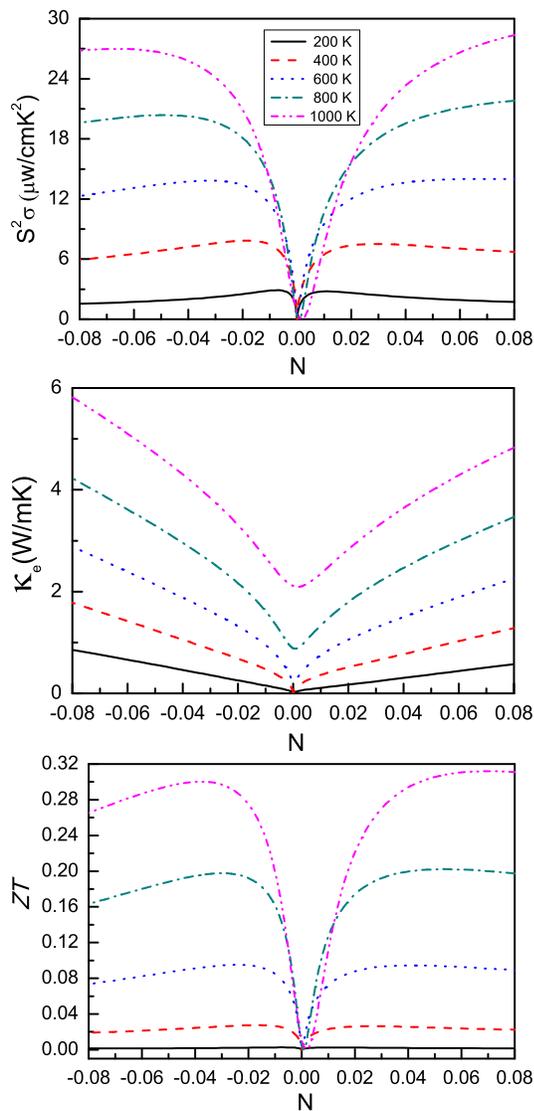}
  \caption{(Color online) The power factor, electronic thermal conductivity and $ZT$ as a function of doping levels with temperature  being 200, 400, 600 , 800 and 1000 (unit: K), and the scattering time $\mathrm{\tau}$  is 0.333 $\times$ $10^{-14}$ s.}\label{s2}
\end{figure}

Here, we assume that the relaxation time $\tau$ is temperature and doping level independent.
The  power factor  $\mathrm{S^2\sigma}$, lattice thermal conductivity  $\mathrm{\kappa_L}$, total thermal conductivity $\mathrm{\kappa}$=$\mathrm{\kappa_e}$+$\mathrm{\kappa_L}$ and $ZT$ as a function of temperature with the doping concentration of $\mathrm{5\times10^{20}cm^{-3}}$ (The doping concentration equals  $\mathrm{1.625\times10^{22}cm^{-3}}$ $\times$ doping level.) for both p-type and n-type are shown in \autoref{s1}. In low temperature region, there is little difference between n-type power factor and p-type one.  At T=1000 K, the power factor is as high as 24.3 $\mathrm{\mu W cm^{-1} K^{-2}}$ for n-type doping, and 20.7 $\mathrm{\mu W cm^{-1} K^{-2}}$ for p-type doping.
The lattice thermal conductivity can be assumed  to be  independent of  doping levels, and
typically goes as 1/T. At room temperature,  the corresponding lattice thermal conductivity is 14.5  $\mathrm{W m^{-1} K^{-1}}$, which is among the reported room temperature lattice thermal conductivity of half-Heuslers (6.7-20 $\mathrm{W m^{-1} K^{-1}}$)\cite{s16}.
At low T, the total thermal conductivity $\mathrm{\kappa}$ is dominated by the $\mathrm{\kappa_L}$, but lattice thermal conductivity $\mathrm{\kappa_L}$ and electronic thermal conductivity $\mathrm{\kappa_e}$ have the same order of magnitude at high
temperature region. At T=1000 K, the $ZT$ is as high as 0.30 for n-type doping, and 0.27 for p-type doping. The high lattice thermal conductivity leads to relatively small $ZT$ value. Finally, the power factor $\mathrm{S^2\sigma}$, electronic thermal conductivity $\mathrm{\kappa_e}$ and $ZT$ as a function of doping levels with temperature  being 200, 400, 600 , 800 and 1000 (unit: K) are plotted in \autoref{s2}. It is found that electronic thermal conductivity increases with the increasing temperature. Calculated results show that the higher $ZT$ can be attained in high doping levels at  high temperature region.

\section{Discussions and Conclusion}
For thermoelectric device applications, although half-Heusler $\mathrm{ZrNiPb}$ has high power factor, high lattice thermal conductivity (14.5  $\mathrm{W m^{-1} K^{-1}}$ at room temperature) is a major disadvantage.  It has been shown that  lattice thermal conductivity can be reduced  by  point defects and boundaries.
The binary  or ternary alloying of Hf, Zr and Ti  in M site of n-type MNiSn and p-type
MCoSb (M=Hf, Zr and Ti) can reduce the thermal conductivity compared with
 a single  element MNiSn\cite{bc1,bc2,bc3,bc4}. The smaller  grain size can  effectively suppress the thermal conductivity.
 The thermal conductivity of $\mathrm{TiNiSn_{1-x}Sb_x}$ is about 3.7  $\mathrm{W m^{-1} K^{-1}}$ with an average
grain size of 50 nm\cite{bc5}, and the room temperature
thermal conductivity is also  3.7  $\mathrm{W m^{-1} K^{-1}}$ at grain sizes below
200 nm for p-type $\mathrm{Hf_{0.5}Zr_{0.5}CoSn_{0.8}Sb_{0.2}}$\cite{bc6}. Therefore, experimentally, it is possible to reduce the thermal conductivity of $\mathrm{ZrNiPb}$ by point defects (doping $\mathrm{ZrNiPb}$ with Hf) and boundaries (reducing grain size).

In summary,  we investigate thermoelectric properties  of  $\mathrm{ZrNiPb}$  based mainly on the reliable first-principle calculations, including  not only electronic  but also lattice part. The electronic part, including power factor and electronic thermal conductivity, is calculated by using GGA+SOC, while lattice thermal conductivity is performed with GGA.
It is found that SOC is very important for p-type power factor calculations, and has a obvious detrimental influence. Calculated results show that  $\mathrm{ZrNiPb}$ has high power factor, but high thermal conductivity is a fatal disadvantage to gain high $ZT$ value.  So, reducing the lattice thermal conductivity is very key to attain better thermoelectric performance.
The present work is useful for further improving thermoelectric performance of half-Heusler $\mathrm{ZrNiPb}$.

\begin{acknowledgments}
This work is  supported by the Fundamental Research Funds for the Central Universities (2015XKMS073). We are grateful to the Advanced Analysis and Computation Center of CUMT for the award of CPU hours to accomplish this work.
\end{acknowledgments}

\end{document}